\documentstyle[twoside,aps]{revtex}

\begin{document}
\draft
\title{A Variational Expansion for the Free Energy of a Bosonic System}
\author{Wen-Fa Lu$^{a,b,c}$,Sang Koo You$^d$, Jino Bak$^{a,c}$,
        Chul Koo Kim$^{a,c}$ and Kyun Nahm$^e$}
\address{$^a$ Institute of Physics and Applied Physics, Yonsei University,
Seoul 120-749, Korea \ \ \\
 $^b$ Department of Physics, and Institute for Theorectical Physics,
     \ \ \ \ \ \ \ \ \ \ \ \ \ \ \ \ \ \ \ \ \ \ \ \ \ \ \ \ \\
         Shanghai Jiao Tong University,
         Shanghai 200030,
        The People's Republic of China
   \thanks{permanent address,E-mail: wenfalu@online.sh.cn}
      \\
 $^c$ Center for Strongly Correlated Materials Research,
Seoul National University, \ \ \ \ \ \ \ \ \ \ \ \\ Seoul 151-742, Korea
 \ \ \ \ \ \ \ \ \ \ \ \ \ \ \ \ \ \ \ \ \ \ \ \ \ \ \ \ \ \ \ \ \ \ \ \ \ \
 \ \ \ \ \ \ \ \ \ \ \ \ \ \ \ \ \ \ \ \ \ \ \ \ \ \
 \ \ \ \ \ \ \ \ \ \ \ \ \ \\
 $^d$ Max-Planck Institute for the Physics of Complex Systems,
      D-01187 Dreaden, Germany  \\
 $^e$ Department of Physics, Yonsei University, Wonju 220-710, Korea
  \ \ \ \ \ \ \ \ \ \ \ \ \ \ \ \ \ \ \ \ \ \ \ \ \
    }
\maketitle

\begin{abstract}
In this paper, a variational perturbation scheme for nonrelativistic
many-Fermion systems is generalized to a Bosonic system. By calculating the
free energy of an anharmonic oscillator model, we investigated
this variational expansion scheme for its efficiency. Using the modified
Feynman rules for the diagrams, we obtained the analytical expression of the
free energy up to the fourth order. Our numerical results at various orders
are compared with the exact and other relevant results.
\end{abstract}


\section{Introduction}
\label{1}

For calculating the free energy of a system, there exist usually two basic
methods, the conventional perturbative and the variational methods \cite{1}.
In fact, these two methods are standard in calculations of many physics
problems. However, it is well known that the former is useful only for small
perturbing potentials, whereas the latter lacks systematic schemes to control
its accuracy, albeit it is valid for any potential. In order to overcome
these difficulties and improve the variational method, a variational
perturbation idea of properly combining the two methods was pioneered by Koehler
in lattice dynamics \cite{2} and Seznec and Zinn-Justin \cite{3} on an
anharmonic osillator decades ago. Later, the idea was extended to path
integrals by Feynman and Kleinert \cite{4} as well as Okopi$\acute{n}$ska
\cite{4}, and further developed by other
authors \cite{5,6,7} \footnote{Here we are far from exausting the relevant
literature.}. Very recently, three of the authors (You, Kim and Nahm) and
their collaborator presented a variational perturbation scheme for a
nonrelativistic many-Fermion system in the functional integral formalism
\cite{8}. In order to test the efficiency of the method, Ref.~\cite{8}
provided a model numerical calculation at the zero temperature up to the
second order. Obviously, this test is limited, and a more wide investigation
is necessary for the variational perturbation scheme.

Actually, the scheme in Ref.~\cite{8} is a Taylor series expansion based on
the variational result obtained in the spirit of Feynman variational principle
\cite{1}. This scheme can improve the variational method if used judicially.
We note that there exists no work performing the same scheme for Bosonic
systems. Although Okopi$\acute{n}$ska \cite{9} and Krzyweck \cite{6}
established two kinds of expansion schemes, the optimized expansions and the
cumulant expansions respectively, whose lowest order results are the
variational result, they are really not based on the variational result
because the variational procedure was performed at the truncated order.
Further, improvement to the variational method is notoriously difficult,
and different schemes will have their own advantages over others. So it is
worthwhile generalizing the scheme in Ref.~\cite{8} to Bosonic systems.

For the above two purposes, an anharmonic oscillator will be an
effective laboratory. For a one-dimensional anharmonic oscillator, the
Hamiltonian is
\begin{equation}
       H(t)={\frac {1}{2 m}} p^2 +{\frac {1}{2}} m\omega^2
            x^2(t) + \lambda x^4(t) \;.
\end{equation}
Here, $x$ is the space position, $p$ the momentum, $t$ the time, and $m,
\omega$ as well as $\lambda$ are the mass, frequency and coupling strength,
respectively. Such an anharmonic oscillator is probably the simplest model
which does not have exact analytic solution. Its exact free energy was
obtained numerically \cite{10,6}. Moreover, there exist many approximation
methods to calculate its free energy \cite{4,6,9,11}. All of these make Eq.(1)
an ideal candidate for our aim. Therefore, in this paper, we will generalize
the scheme in Ref.~\cite{8}, taking the simple anharmonic oscillator, Eq.(1),
as a laboratory and investigate the Taylor expansion scheme on the free energy
based on the variational result obtained from the Feynman variational
principle \cite{1}.

Simultaneously, the anharmonic oscillator itself is useful in chemical physics
\cite{12} and many physics problems, such as thermal expansion, phonon
softening and structural phase transitions \cite{13}. Although its free energy
was calculated by using many methods, their numerical (approximate and exact)
results \footnote{In this paper, an anharmonic oscillator does not include the
double-well potential case.} were focused mainly on the case of the reduced
temperature $T$ less than 1, except for Ref.~\cite{10} which provided
accurate results at moderate $T$. Therefore, we also test whether our scheme
can produce reliable results at moderate temperature range.

We will work within the functional integral formalism \cite{14}. Next section
generalizes the scheme in Ref.~\cite{8} to calculate the free energy of a
system with the potential $V(x)$. In Sect.III, we carry out the scheme on the
anharmonic oscillator, Eq.(1). Sect.IV presents calculations of the free
energy for Eq.(1) up to the fourth order and makes comparisons with the exact,
and various approximate results, such as, variational, cumulant-expansion and
optimized-expansion results. Conclusions are given in Sect.V.

\section{A Variational Expasion of Free energies for Bosonic Systems}
\label{2}

For a system with the Hamiltonian $H(t)={\frac {1}{2 m}} p^2
+V[x(t)]$ with $V[x(t)]={\frac {1}{2}} m\omega^2 x^2(t) + V_I[x(t)]$, the
generating functional is \cite{14} (1990)
\begin{equation}
Z[J]= \int_{x(0)=x(\beta)} {\cal D}[x(\tau)] \exp\{
      - \int_{0}^{\beta}[{\frac {1}{2}}m x(\tau)(-\partial_\tau^2)
      x(\tau) + V[x(\tau)] - J(\tau)x(\tau)]d\tau\}  \;,
\end{equation}
where $\tau=it$ is the imaginary time, $\partial_\tau \equiv {\frac
{\partial}{\partial \tau}}$, $\beta={\frac {1}{\kappa T}}$ with
$\kappa$ the Boltzmann constant (we will consider it as unity for convenience)
and $T$ the temperature. $J$ is an arbitrary external source and the symbol
${\cal D}[x(\tau)]$ represents the functional measure. In Eq.(2), $x(0)=
x(\beta)$ means that the functional integral should be executed over all the
closed paths \cite{14} (1990). To introduce a variational parameter $\Omega$,
one can rewrite the classical action functional $S[x]$ in the Euclidean
space-time in Eq.(2) as \cite{3,6,9}
\begin{eqnarray}
S[x,J]&=&\int_{0}^{\beta} [{\frac {1}{2}}m x(\tau)(-\partial_\tau^2+\Omega^2)
      x(\tau) - J(\tau)x(\tau) - {\frac {1}{2}} m\Omega^2 x^2(\tau)
      + V[x(\tau)]]d\tau    \nonumber \\
      & \ & \equiv S_0[x,J]+S_D[x]
\end{eqnarray}
with $S_0[x,J]=\int_{0}^{\beta} [{\frac {1}{2}}m x(\tau)(-\partial_\tau^2
+\Omega^2)x(\tau) - J(\tau)x(\tau)]d\tau$ and $S_D[x]=\int_{0}^{\beta}
[- {\frac {1}{2}} m\Omega^2 x^2(\tau) + V[x(\tau)]]d\tau$.
Thus, $Z[J]$ can be rewritten as
\begin{eqnarray}
Z[J]&=& \int_{x(0)=x(\beta)} {\cal D}[x(\tau)] \exp\{-S_0[x,J]-S_D[x]\}
        \nonumber \\
    &=& \exp\{- \int_{0}^{\beta}[-{\frac {1}{2}}m\Omega^2\delta^2
      _{J_{\tau}} + V[\delta_{J_\tau}]]d\tau\}
      \int_{x(0)=x(\beta)} {\cal D}[x(\tau)] \exp\{-S_0[x,J]\}
       \nonumber  \\
    &=& (Det(G^{-1}))^{-{\frac {1}{2}}}
        \exp\{- \int_{0}^{\beta}[-{\frac {1}{2}}m(\Omega^2-\omega^2)\delta^2
      _{J_{\tau}} + V_I [\delta_{J_\tau}]]d\tau\}
        \exp\{{\frac {1}{2}}J_\tau G_{\tau\tau'}J_{\tau'}
           d\tau d\tau'\} \;,
\end{eqnarray}
where, $J_\tau=J(\tau)$, $\delta_{J_\tau}\equiv {\frac {\delta}
{\delta J_\tau}}$, $G^{-1}$ represents the operator $-\partial^2_\tau +
\Omega^2$ with the propagator $G_{\tau\tau'}=G(\tau,\tau')$ and $Det$ means
the determinant. In the third equation of Eq.(4), we have carried out the
Gaussian functional integration \cite{14} (1990). Note that $G_{\tau\tau'}$
can be expanded into a series owing to the closed path requirement in the
functional integral, Eq.(2) \cite{14,15}. To calculate the free energy, we
express the partition function $Z\equiv Z[J=0]$ in the following form,
\begin{eqnarray}
Z&=& \int_{x(0)=x(\beta)} {\cal D}[x(\tau)] \exp\{-S_0[x,J=0]\}
     {\frac {\int_{x(0)=x(\beta)} {\cal D}[x(\tau)]
     \exp\{-S_0[x,J=0]-S_D[x]\}}{\int_{x(0)=x(\beta)} {\cal D}[x(\tau)]
     \exp\{-S_0[x,J=0]\}}}
        \nonumber \\
 &=& (Det(G^{-1}))^{-{\frac {1}{2}}}<\exp\{
      - \int_{0}^{\beta}[-{\frac {1}{2}}m(\Omega^2-\omega^2)\delta^2
      _{J_{\tau}} + V_I [\delta_{J_\tau}]]d\tau\}>_G  \\
 &=& (Det(G^{-1}))^{-{\frac {1}{2}}} \exp\{
      - \int_{0}^{\beta}<-{\frac {1}{2}}m(\Omega^2-\omega^2)\delta^2
      _{J_{\tau}} + V_I [\delta_{J_\tau}]>_G d\tau\}  \nonumber \\
    && \cdot  <\exp\{
      - \int_{0}^{\beta}[-{\frac {1}{2}}m(\Omega^2-\omega^2)\delta^2
      _{J_{\tau}} + V_I [\delta_{J_\tau}] \nonumber \\  && \ \ \ \ \
    - <-{\frac {1}{2}}m(\Omega^2-\omega^2)\delta^2
      _{J_{\tau}} + V_I [\delta_{J_\tau}]>_G]d\tau\}>_G  \;.
\end{eqnarray}
Here, we have used the following notation and relation
\begin{eqnarray}
<O[x]>_G & \equiv & {\frac {\int_{x(0)=x(\beta)} {\cal D}[x(\tau)]
             O[x]\exp\{
      - \int_{0}^{\beta}{\frac {1}{2}}m x(\tau)(-\partial^2_\tau + \Omega^2)
      x(\tau)d\tau\}}{\int_{x(0)=x(\beta)} {\cal D}[x(\tau)]
      \exp\{- \int_{0}^{\beta}{\frac {1}{2}}m x(\tau)(-\partial^2_\tau +
      \Omega^2)x(\tau)d\tau\}}} \nonumber \\
      &=& O[\delta_J] \exp\{{\frac {1}{2}}J_\tau G_{\tau\tau'}J_{\tau'}
           d\tau d\tau'\}\bigg|_{J=0} \equiv <O[\delta_J]>_G  \; .
\end{eqnarray}
Obviously, when $V_I[x(t)]$ is not zero, it will be impossible to obtain
analytically exact partition function and, hence, one has to design some
scheme to produce an approximate solution. For the case of small $V_I$, one
can take $\Omega=\omega$, make a Taylor series expansion to the exponential in
Eq.(5) and then truncate the series at some order to approximate $Z$. This is
just the conventional perturbation theory and the propergator $G_{\tau\tau'}$
is the bare propagator. When $V_I$ is not so small, the above perturbation
method is valid no longer. In such a case, an effective alternative is the
variational method which is based on the Fenyman variational principle. In the
following, we briefly introduce the variational method.

Exploiting the Jensen's inequality \cite{1,8}, one can have
\begin{eqnarray}
 &&<\exp\{- \int_{0}^{\beta}[-{\frac {1}{2}}m(\Omega^2-\omega^2)\delta^2
      _{J_{\tau}} + V_I [\delta_{J_\tau}]]d\tau\}>_G
      \nonumber \\
  &\ge&  \exp\{- \int_{0}^{\beta}<-{\frac {1}{2}}m(\Omega^2-\omega^2)\delta^2
      _{J_{\tau}} + V_I [\delta_{J_\tau}]>_G d\tau\}  \;.
\end{eqnarray}
Substituting the above equation into Eq.(6) leads to a relation for the lower
limit of the partition function, $i.e.$,
\begin{equation}
Z \ge (Det(G^{-1}))^{-{\frac {1}{2}}} \exp\{
      - \int_{0}^{\beta}<-{\frac {1}{2}}m(\Omega^2-\omega^2)\delta^2
      _{J_{\tau}} + V_I [\delta_{J_\tau}]>_G d\tau\}  \;.
\end{equation}
Hence, the free energy is
\begin{equation}
F=-{\frac {1}{\beta}}\ln(Z)\le {\frac {1}{2\beta}} \ln(Det(G^{-1})) +
   {\frac {1}{\beta}}\int_{0}^{\beta}
   <-{\frac {1}{2}}m(\Omega^2-\omega^2)\delta^2
      _{J_{\tau}} + V_I [\delta_{J_\tau}]>_G d\tau \equiv {\bar F} \;.
\end{equation}
Obviously, making ${\bar F}$ of the last equation the absolute minimum will
lead to a minimum upper limit of the free energy, $F_0$, with the
variationally extremized condition,
\begin{equation}
{\frac {\delta {\bar F}}
{\delta \Omega^2}}=0   \;,
\end{equation}
and the stabilized condition
\begin{equation}
{\frac {\delta^2 {\bar F}}
{(\delta \Omega^2)^2}}\ge 0  \;.
\end{equation}
The parameter $\Omega$  which renders ${\bar F}$ absolutely minimized will be
chosen from the three possibilities: the non-zero solution of Eq.(11), zero
and $\infty$. ${\bar F}$ with such an $\Omega$ is just $F_0$, the
variational result of $F$. This procedure is essentially same as was done in
Ref.~\cite{16} for Eq.(1).

Entering the above variational result into Eq.(6) and taking the logarithm,
we obtain the following expression of $F$:
\begin{equation}
F = F_0 -{\frac {1}{\beta}}\ln[<\exp\{-[S_D[\delta_{J}] -
    <S_D[\delta_{J}]>_G]\}>_G]  \;.
\end{equation}
Now, we make a Taylor series expansion of the exponential in the last
equation, and the average $<\cdots>_G$ can be calculated order by order
through borrowing the Feynman diagram technique \cite{14} \footnote{Now
diagrams are no longer the bare Feynman diagrams in the sense of perturbation
theory owing to $\Omega \not=\omega$ which has been variationally determined.}.
The logarithmic operation in Eq.(13) is equivalent to discarding disconnected
diagrams \cite{14}. Consequently, we have
\begin{equation}
F = F_0 + \sum^\infty_2 F^{(n)}
\end{equation}
with the $n$th order correction to the variational result
\begin{eqnarray}
F^{(n)} &=& (-1)^{n+1} {\frac {1}{\beta}} {\frac {1}{n!}}
    <[S_D[\delta_J]|_{\tau=\tau_1} - < S_D[\delta_J]|_{\tau=\tau_1}>_G]
           \nonumber \\  && \cdots
    [S_D[\delta_J]|_{\tau=\tau_i} - < S_D[\delta_J]|_{\tau=\tau_i}>_G]
         \nonumber  \\  && \cdots
   [S_D[\delta_J]|_{\tau=\tau_n} - < S_D[\delta_J]|_{\tau=\tau_n}>_G]
      >_{G,C}   \; .
\end{eqnarray}
Here, the subscript $C$ means that only the connected diagrams have their
contributions to the free energy \cite{14}. Eq.(14) corresponds to a
systematic Feynman-diagram-like expansion, and from it, one can estimate the
approximate values of the free energy, $F$, order by order.
Thus, we have finished the generalization of the scheme in Ref.~\cite{8} to
a Bosonic case. In this scheme, the parameter $\Omega$
is variationally determined before the series expansion is performed, and
it is identical for all orders. Obviously, this is a Taylor series expansion
around the variational result, and so we call it the variational expansion.
Because the $i$th factor in Eq.(15) ($i=1,2, \cdots n$) has the term $-<S_D
[\delta_J]|_{\tau=\tau_i}>_G= - \int_{0}^{\beta}<-{\frac {1}{2}}m\Omega^2
\delta^2_{J_{\tau_i}} + V[\delta_{J_{\tau_i}}]>_G d\tau_i$, which has negative
sign against the major part of $F_0$, one can expect to get a simplified
diagram rule, as will be shown for the system, Eq.(1), in the next section.
Next, we apply the above procedure to the anharmonic oscillator and make a
comparison with existing results in the literature.

\section{Application to the anharmonic oscillator}
\label{3}

For the system, Eq.(1) ($V[x(t)]={\frac {1}{2}} m\omega^2 x^2(t) + \lambda
x^4(t)$), the procedure from Eq.(8) to Eq.(12) yields easily the variational
free energy, $F_0$,
\begin{equation}
F_0={\frac {1}{\beta}}\ln(2 \sinh({\frac {\beta \Omega}{2}}))
    -{\frac {3 \lambda}{4 m^2\Omega^2}}\coth^2({\frac {\beta \Omega}{2}})
\end{equation}
with the variationally extremized condition (${\frac {\delta {\bar F}}
{\delta \Omega^2}}=0$)
\begin{equation}
\Omega^2=\omega^2 + {\frac {6 \lambda}{m^2\Omega}}\coth
                 ({\frac {\beta \Omega}{2}}) \; .
\end{equation}
Here, owing to the periodicity of the path in Eq.(2), we have used the
following propagator
\begin{equation}
G_{\tau\tau'}={\frac {1}{\beta}} \sum_{-\infty}^{\infty}
     {\frac {1}{m (\omega_n^2+\Omega^2)}}e^{-i\omega_n (\tau-\tau')}
     ={\frac {1}{2 m \Omega}}{\frac {\cosh({\frac {\beta \Omega}{2}}-
       \Omega |\tau-\tau'|)}{\sinh({\frac {\beta \Omega}{2}})}}
\end{equation}
with $\omega_n$ the Matsubara frequency \cite{14,15}. Eq.(16) coupled with
Eq.(17) are just the variational result of $F$ in Ref.~\cite{16}.

Using the relation ${\frac {1}{2}}m^2 (\omega^2-\Omega^2)=-6\lambda
G_{\tau\tau}$ from Eq.(17), we have
\begin{eqnarray}
<S_D[\delta_J]>_G&=&\int_{0}^{\beta}<{\frac {1}{2}}m(\omega^2
-\Omega^2)\delta^2_{J_{\tau}} + \lambda\delta^4_{J_{\tau}}>_G d\tau
              \nonumber  \\
   & = & \int_{0}^{\beta}[{\frac {1}{2}}m(\omega^2-\Omega^2)G_{\tau\tau}
         + 3\lambda G_{\tau\tau}G_{\tau\tau}]d\tau= -
   \int_{0}^{\beta} 3\lambda G_{\tau\tau}G_{\tau\tau} d\tau
         \;.
\end{eqnarray}
So, for any $i$, one has
\begin{eqnarray}
&&<\cdots S_D[\delta_J]|_{\tau=\tau_i} - < S_D[\delta_J]|_{\tau=\tau_i}>_G]
   \cdots >_{G,C} \nonumber  \\
   & = & <\cdots\int_{0}^{\beta}[{\frac {1}{2}}m(\omega^2
   -\Omega^2)\delta^2_{J_{\tau_i}} + \lambda\delta^4_{J_{\tau_i}}-
   <-{\frac {1}{2}}m(\omega^2-\Omega^2)\delta^2_{J_{\tau_i}} +
   \lambda\delta^4_{J_{\tau_i}}>_G d\tau_i\cdots>_{G,C}  \nonumber  \\
   &=& <\cdots\int_{0}^{\beta}[{\frac {1}{2}}m(\omega^2
   -\Omega^2)((\stackrel{\circ}{\delta}_{J_{\tau_i}})^2
   + G_{\tau_i\tau_i})   \nonumber  \\ & \ & \ \ \ \ \ \ \ \ \
   + \lambda((\stackrel{\circ}{\delta}_
   {J_{\tau_i}})^4 + 6 G_{\tau_i\tau_i}(\stackrel{\circ}{\delta}_
   {J_{\tau_i}})^2+ 3G_{\tau_i\tau_i}G_{\tau_i\tau_i})
   + 3\lambda G_{\tau_i\tau_i}G_{\tau_i\tau_i}]d\tau_i\cdots>_{G,C}
   \nonumber   \\
   & = & <\cdots\int_{0}^{\beta}[\lambda((\stackrel{\circ}{\delta}_
   {J_{\tau_i}})^4 \cdots>_{G,C}  \;,
\end{eqnarray}
where, the symbol $\circ$ in $\stackrel{\circ}{\delta}$ means that the
functional derivative with the index $i$ takes effects on
$\exp\{{\frac {1}{2}}J_\tau G_{\tau\tau'}J_{\tau'}d\tau d\tau'\}$
only if it makes up pair with any other functional derivative with the
index $j\not= i$ to yield $G_{\tau_i\tau_j}$. 
In going to the second step of the last equation, a concrete analysis has led
to the following equivalent properties ($\Leftrightarrow$ means equivalence):
\begin{equation}
\delta^2_{J_{\tau_i}}\Leftrightarrow (\stackrel{\circ}{\delta}_{J_{\tau_i}})^2
   + G_{\tau_i\tau_i} \;, \ \ \ \ \
   \delta^4_{J_{\tau_i}}\Leftrightarrow (\stackrel{\circ}{\delta}_
   {J_{\tau_i}})^4 + 6 G_{\tau_i\tau_i}(\stackrel{\circ}{\delta}_
   {J_{\tau_i}})^2+ 3 G_{\tau_i\tau_i}G_{\tau_i\tau_i} \;.
\end{equation}
In terms of the Feynman diagram language, it implies that only the legs which
come from different vertices can connect each other.

Substituting Eq.(20) into Eq.(14), we can estimate the higher-order
corrections to $F_0$ in Eq.(16) with the help of the diagram technique
\cite{14}. The free energy for Eq.(1) is now
\begin{equation}
F= F_0 -{\frac {1}{\beta}}<\exp\{-\int_0^\beta\lambda(\stackrel
   {\circ}{\delta}_{J_\tau})^4 d\tau\}>_{G,C}=F_0 + \sum^\infty_2 F^{(n)}
\end{equation}
with the $n$th order correction,
\begin{eqnarray}
F^{(n)} = (-1)^{n+1} {\frac {1}{\beta}} {\frac {\lambda^n}{n!}}
    <\int_0^\beta d\tau_1(\stackrel{\circ}{\delta}_{J_{\tau_1}})^4\cdots
    \int_0^\beta d\tau_n(\stackrel{\circ}{\delta}_{J_{\tau_n}})^4>_{G,C} \; .
\end{eqnarray}
Here, the modified Feynman's rules for drawing diagrams are
quite simple, and they are as follows: \\
\unitlength=1mm
\begin{picture}(150,14)(0,0)
\put(5,6){(1).}
\put(15,6){\line(1,0){15}}
\put(15,6){\circle*{0.6}}
\put(30,6){\circle*{0.6}}
\put(35,6){Propagator, \ \ \ \ \ $G_{\tau_1\tau_2}$ ;}
\put(93,6){(2).}
\put(103,11){\circle*{0.6}}
\put(103,1){\circle*{0.6}}
\put(108,6){\circle*{0.6}}
\put(108,6){\line(1,-1){5}}
\put(108,6){\line(-1,-1){5}}
\put(108,6){\line(1,1){5}}
\put(108,6){\line(-1,1){5}}
\put(113,11){\circle*{0.6}}
\put(113,1){\circle*{0.6}}
\put(118,6){Vertex, \ \ \ \ \ $-\lambda\int_0^\beta d\tau$ .}
\end{picture}    \\
For the $n$th order, there is an additional total factor $-{\frac {1}
{\beta n!}}$ . From Eq.(23), it is evident that there will be not any Cactus
diagrams appearing at any higher order, which is demonstrated by the diagrams
in the next section. This simplifying feature of diagrams is similar to what
occurs in the Fermionic case \cite{8}. A further analysis indicates that
there exist the following four types of building bricks for any $n$th-order
connected diagrams ($n>2$): \\
\unitlength=1mm
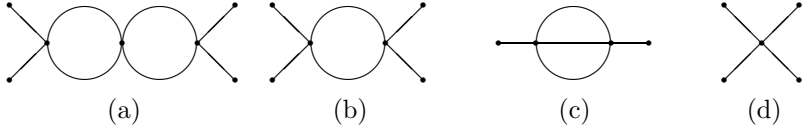
\begin{picture}(150,24)(0,0)
\put(20,20){\circle*{0.6}}
\put(20,10){\circle*{0.6}}
\put(25,15){\circle*{0.6}}
\put(25,15){\line(-1,1){5}}
\put(25,15){\line(-1,-1){5}}
\put(30,15){\circle{10}}
\put(35,15){\circle*{0.6}}
\put(33,5){(a)}
\put(40,15){\circle{10}}
\put(45,15){\circle*{0.6}}
\put(45,15){\line(1,1){5}}
\put(45,15){\line(1,-1){5}}
\put(50,20){\circle*{0.6}}
\put(50,10){\circle*{0.6}}
\put(55,20){\circle*{0.6}}
\put(55,10){\circle*{0.6}}
\put(60,15){\line(-1,1){5}}
\put(60,15){\line(-1,-1){5}}
\put(60,15){\circle*{0.6}}
\put(65,15){\circle{10}}
\put(63,5){(b)}
\put(70,15){\circle*{0.6}}
\put(70,15){\line(1,1){5}}
\put(70,15){\line(1,-1){5}}
\put(75,20){\circle*{0.6}}
\put(75,10){\circle*{0.6}}
\put(85,15){\circle*{0.6}}
\put(90,15){\line(-1,0){5}}
\put(90,15){\circle*{0.6}}
\put(90,15){\line(1,0){5}}
\put(95,15){\circle{10}}
\put(93,5){(c)}
\put(100,15){\line(-1,0){5}}
\put(100,15){\circle*{0.6}}
\put(100,15){\line(1,0){5}}
\put(105,15){\circle*{0.6}}
\put(115,20){\circle*{0.6}}
\put(115,10){\circle*{0.6}}
\put(120,15){\line(-1,1){5}}
\put(120,15){\line(-1,-1){5}}
\put(120,15){\circle*{0.6}}
\put(118,5){(d)}
\put(120,15){\line(1,1){5}}
\put(120,15){\line(1,-1){5}}
\put(125,20){\circle*{0.6}}
\put(125,10){\circle*{0.6}}
\end{picture}  \\
which correspond to the four kinds of partitionings of the integer ``4'' :
(a) 2+2 , \ (b) 2+1+1 , \ (c) 3+1 and (d) 1+1+1+1, respectively. In this
figure, the intermediate vertex of the brick (a) has two
legs connected with one different vertex and the other two legs with another
different vertex; the left (or right) vertex of the brick (b) has two legs
connected with one different vertex and the other two legs of it will be
connected with some other two different vertices respectively; the left (or
right) vertex of the brick (c) has three legs connected with one different
vertex and the other leg of it will be connected with some different vertex;
and the vertex of the brick (d) will have its legs connected with some four
different vertices respectively. These four bricks are helpful for drawing
various distinct diagrams at any order as one can see from the five
diagrams drawn in the next section. For example, all of them do not contain
the brick (d), the second-order diagram consists of only the brick (a), so
does the first diagram of the forth-order diagrams (4a).

In the next section, we calculate the free energy up to the fourth order from
Eq.(22).

\section{Analytical Expressions and Numerical Results \\
          up to the Fourth order}
\label{4}

According to the last section, the topologically non-equivalent diagrams at
the second, third and fourth orders can be drawn as follows,

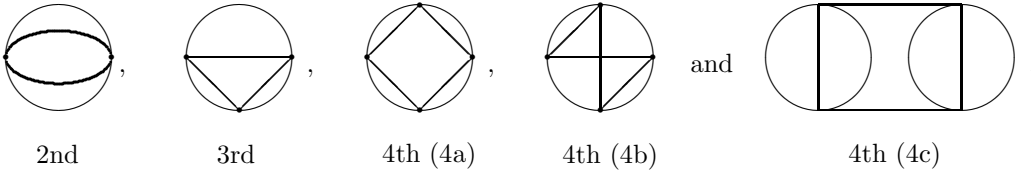
\unitlength=1mm
\begin{picture}(150,30)(0,0)
\put(10,20){\circle*{0.6}}
\put(17,20){\circle{14}}
\put(24,20){\circle*{0.6}}
\put(10.00,20.00){\circle*{0.05}}   \put(10.00,20.00){\circle*{0.1}}
\put(10.10,20.59){\circle*{0.05}}   \put(10.10,19.41){\circle*{0.1}}
\put(10.20,20.83){\circle*{0.05}}   \put(10.20,19.17){\circle*{0.1}}
\put(10.30,21.01){\circle*{0.05}}   \put(10.30,18.99){\circle*{0.1}}
\put(10.40,21.17){\circle*{0.05}}   \put(10.40,18.83){\circle*{0.1}}
\put(10.50,21.30){\circle*{0.05}}   \put(10.50,18.70){\circle*{0.1}}
\put(10.60,21.42){\circle*{0.05}}   \put(10.60,18.58){\circle*{0.1}}
\put(10.70,21.53){\circle*{0.05}}   \put(10.70,18.47){\circle*{0.1}}
\put(10.80,21.62){\circle*{0.05}}   \put(10.80,18.38){\circle*{0.1}}
\put(10.90,21.72){\circle*{0.05}}   \put(10.90,18.28){\circle*{0.1}}
\put(11.00,21.80){\circle*{0.05}}   \put(11.00,18.20){\circle*{0.1}}
\put(11.10,21.88){\circle*{0.05}}   \put(11.10,18.12){\circle*{0.1}}
\put(11.20,21.96){\circle*{0.05}}   \put(11.20,18.04){\circle*{0.1}}
\put(11.30,22.03){\circle*{0.05}}   \put(11.30,17.97){\circle*{0.1}}
\put(11.40,22.10){\circle*{0.05}}   \put(11.40,17.90){\circle*{0.1}}
\put(11.50,22.17){\circle*{0.05}}   \put(11.50,17.83){\circle*{0.1}}
\put(11.60,22.23){\circle*{0.05}}   \put(11.60,17.77){\circle*{0.1}}
\put(11.70,22.29){\circle*{0.05}}   \put(11.70,17.71){\circle*{0.1}}
\put(11.80,22.34){\circle*{0.05}}   \put(11.80,17.66){\circle*{0.1}}
\put(11.90,22.40){\circle*{0.05}}   \put(11.90,17.60){\circle*{0.1}}
\put(12.00,22.45){\circle*{0.05}}   \put(12.00,17.55){\circle*{0.1}}
\put(12.10,22.50){\circle*{0.05}}   \put(12.10,17.50){\circle*{0.1}}
\put(12.20,22.55){\circle*{0.05}}   \put(12.20,17.45){\circle*{0.1}}
\put(12.30,22.59){\circle*{0.05}}   \put(12.30,17.41){\circle*{0.1}}
\put(12.40,22.64){\circle*{0.05}}   \put(12.40,17.36){\circle*{0.1}}
\put(12.50,22.68){\circle*{0.05}}   \put(12.50,17.32){\circle*{0.1}}
\put(12.60,22.72){\circle*{0.05}}   \put(12.60,17.28){\circle*{0.1}}
\put(12.70,22.76){\circle*{0.05}}   \put(12.70,17.24){\circle*{0.1}}
\put(12.80,22.80){\circle*{0.05}}   \put(12.80,17.20){\circle*{0.1}}
\put(12.90,22.84){\circle*{0.05}}   \put(12.90,17.16){\circle*{0.1}}
\put(13.00,22.87){\circle*{0.05}}   \put(13.00,17.13){\circle*{0.1}}
\put(13.10,22.91){\circle*{0.05}}   \put(13.10,17.09){\circle*{0.1}}
\put(13.20,22.94){\circle*{0.05}}   \put(13.20,17.06){\circle*{0.1}}
\put(13.30,22.97){\circle*{0.05}}   \put(13.30,17.03){\circle*{0.1}}
\put(13.40,23.00){\circle*{0.05}}   \put(13.40,17.00){\circle*{0.1}}
\put(13.50,23.03){\circle*{0.05}}   \put(13.50,16.97){\circle*{0.1}}
\put(13.60,23.06){\circle*{0.05}}   \put(13.60,16.94){\circle*{0.1}}
\put(13.70,23.09){\circle*{0.05}}   \put(13.70,16.91){\circle*{0.1}}
\put(13.80,23.11){\circle*{0.05}}   \put(13.80,16.89){\circle*{0.1}}
\put(13.90,23.14){\circle*{0.05}}   \put(13.90,16.86){\circle*{0.1}}
\put(14.00,23.16){\circle*{0.05}}   \put(14.00,16.84){\circle*{0.1}}
\put(14.10,23.19){\circle*{0.05}}   \put(14.10,16.81){\circle*{0.1}}
\put(14.20,23.21){\circle*{0.05}}   \put(14.20,16.79){\circle*{0.1}}
\put(14.30,23.23){\circle*{0.05}}   \put(14.30,16.77){\circle*{0.1}}
\put(14.40,23.25){\circle*{0.05}}   \put(14.40,16.75){\circle*{0.1}}
\put(14.50,23.27){\circle*{0.05}}   \put(14.50,16.73){\circle*{0.1}}
\put(14.60,23.29){\circle*{0.05}}   \put(14.60,16.71){\circle*{0.1}}
\put(14.70,23.31){\circle*{0.05}}   \put(14.70,16.69){\circle*{0.1}}
\put(14.80,23.32){\circle*{0.05}}   \put(14.80,16.68){\circle*{0.1}}
\put(14.90,23.34){\circle*{0.05}}   \put(14.90,16.66){\circle*{0.1}}
\put(15.00,23.35){\circle*{0.05}}   \put(15.00,16.65){\circle*{0.1}}
\put(15.10,23.37){\circle*{0.05}}   \put(15.10,16.63){\circle*{0.1}}
\put(15.20,23.38){\circle*{0.05}}   \put(15.20,16.62){\circle*{0.1}}
\put(15.30,23.40){\circle*{0.05}}   \put(15.30,16.60){\circle*{0.1}}
\put(15.40,23.41){\circle*{0.05}}   \put(15.40,16.59){\circle*{0.1}}
\put(15.50,23.42){\circle*{0.05}}   \put(15.50,16.58){\circle*{0.1}}
\put(15.60,23.43){\circle*{0.05}}   \put(15.60,16.57){\circle*{0.1}}
\put(15.70,23.44){\circle*{0.05}}   \put(15.70,16.56){\circle*{0.1}}
\put(15.80,23.45){\circle*{0.05}}   \put(15.80,16.55){\circle*{0.1}}
\put(15.90,23.46){\circle*{0.05}}   \put(15.90,16.54){\circle*{0.1}}
\put(16.00,23.46){\circle*{0.05}}   \put(16.00,16.54){\circle*{0.1}}
\put(16.10,23.47){\circle*{0.05}}   \put(16.10,16.53){\circle*{0.1}}
\put(16.20,23.48){\circle*{0.05}}   \put(16.20,16.52){\circle*{0.1}}
\put(16.30,23.48){\circle*{0.05}}   \put(16.30,16.52){\circle*{0.1}}
\put(16.40,23.49){\circle*{0.05}}   \put(16.40,16.51){\circle*{0.1}}
\put(16.50,23.49){\circle*{0.05}}   \put(16.50,16.51){\circle*{0.1}}
\put(16.60,23.49){\circle*{0.05}}   \put(16.60,16.51){\circle*{0.1}}
\put(16.70,23.50){\circle*{0.05}}   \put(16.70,16.50){\circle*{0.1}}
\put(16.80,23.50){\circle*{0.05}}   \put(16.80,16.50){\circle*{0.1}}
\put(16.90,23.50){\circle*{0.05}}   \put(16.90,16.50){\circle*{0.1}}
\put(17.00,23.50){\circle*{0.05}}   \put(17.00,16.50){\circle*{0.1}}
\put(17.10,23.50){\circle*{0.05}}   \put(17.10,16.50){\circle*{0.1}}
\put(17.20,23.50){\circle*{0.05}}   \put(17.20,16.50){\circle*{0.1}}
\put(17.30,23.50){\circle*{0.05}}   \put(17.30,16.50){\circle*{0.1}}
\put(17.40,23.49){\circle*{0.05}}   \put(17.40,16.51){\circle*{0.1}}
\put(17.50,23.49){\circle*{0.05}}   \put(17.50,16.51){\circle*{0.1}}
\put(17.60,23.49){\circle*{0.05}}   \put(17.60,16.51){\circle*{0.1}}
\put(17.70,23.48){\circle*{0.05}}   \put(17.70,16.52){\circle*{0.1}}
\put(17.80,23.48){\circle*{0.05}}   \put(17.80,16.52){\circle*{0.1}}
\put(17.90,23.47){\circle*{0.05}}   \put(17.90,16.53){\circle*{0.1}}
\put(18.00,23.46){\circle*{0.05}}   \put(18.00,16.54){\circle*{0.1}}
\put(18.10,23.46){\circle*{0.05}}   \put(18.10,16.54){\circle*{0.1}}
\put(18.20,23.45){\circle*{0.05}}   \put(18.20,16.55){\circle*{0.1}}
\put(18.30,23.44){\circle*{0.05}}   \put(18.30,16.56){\circle*{0.1}}
\put(18.40,23.43){\circle*{0.05}}   \put(18.40,16.57){\circle*{0.1}}
\put(18.50,23.42){\circle*{0.05}}   \put(18.50,16.58){\circle*{0.1}}
\put(18.60,23.41){\circle*{0.05}}   \put(18.60,16.59){\circle*{0.1}}
\put(18.70,23.40){\circle*{0.05}}   \put(18.70,16.60){\circle*{0.1}}
\put(18.80,23.38){\circle*{0.05}}   \put(18.80,16.62){\circle*{0.1}}
\put(18.90,23.37){\circle*{0.05}}   \put(18.90,16.63){\circle*{0.1}}
\put(19.00,23.35){\circle*{0.05}}   \put(19.00,16.65){\circle*{0.1}}
\put(19.10,23.34){\circle*{0.05}}   \put(19.10,16.66){\circle*{0.1}}
\put(19.20,23.32){\circle*{0.05}}   \put(19.20,16.68){\circle*{0.1}}
\put(19.30,23.31){\circle*{0.05}}   \put(19.30,16.69){\circle*{0.1}}
\put(19.40,23.29){\circle*{0.05}}   \put(19.40,16.71){\circle*{0.1}}
\put(19.50,23.27){\circle*{0.05}}   \put(19.50,16.73){\circle*{0.1}}
\put(19.60,23.25){\circle*{0.05}}   \put(19.60,16.75){\circle*{0.1}}
\put(19.70,23.23){\circle*{0.05}}   \put(19.70,16.77){\circle*{0.1}}
\put(19.80,23.21){\circle*{0.05}}   \put(19.80,16.79){\circle*{0.1}}
\put(19.90,23.19){\circle*{0.05}}   \put(19.90,16.81){\circle*{0.1}}
\put(20.00,23.16){\circle*{0.05}}   \put(20.00,16.84){\circle*{0.1}}
\put(20.10,23.14){\circle*{0.05}}   \put(20.10,16.86){\circle*{0.1}}
\put(20.20,23.11){\circle*{0.05}}   \put(20.20,16.89){\circle*{0.1}}
\put(20.30,23.09){\circle*{0.05}}   \put(20.30,16.91){\circle*{0.1}}
\put(20.40,23.06){\circle*{0.05}}   \put(20.40,16.94){\circle*{0.1}}
\put(20.50,23.03){\circle*{0.05}}   \put(20.50,16.97){\circle*{0.1}}
\put(20.60,23.00){\circle*{0.05}}   \put(20.60,17.00){\circle*{0.1}}
\put(20.70,22.97){\circle*{0.05}}   \put(20.70,17.03){\circle*{0.1}}
\put(20.80,22.94){\circle*{0.05}}   \put(20.80,17.06){\circle*{0.1}}
\put(20.90,22.91){\circle*{0.05}}   \put(20.90,17.09){\circle*{0.1}}
\put(21.00,22.87){\circle*{0.05}}   \put(21.00,17.13){\circle*{0.1}}
\put(21.10,22.84){\circle*{0.05}}   \put(21.10,17.16){\circle*{0.1}}
\put(21.20,22.80){\circle*{0.05}}   \put(21.20,17.20){\circle*{0.1}}
\put(21.30,22.76){\circle*{0.05}}   \put(21.30,17.24){\circle*{0.1}}
\put(21.40,22.72){\circle*{0.05}}   \put(21.40,17.28){\circle*{0.1}}
\put(21.50,22.68){\circle*{0.05}}   \put(21.50,17.32){\circle*{0.1}}
\put(21.60,22.64){\circle*{0.05}}   \put(21.60,17.36){\circle*{0.1}}
\put(21.70,22.59){\circle*{0.05}}   \put(21.70,17.41){\circle*{0.1}}
\put(21.80,22.55){\circle*{0.05}}   \put(21.80,17.45){\circle*{0.1}}
\put(21.90,22.50){\circle*{0.05}}   \put(21.90,17.50){\circle*{0.1}}
\put(22.00,22.45){\circle*{0.05}}   \put(22.00,17.55){\circle*{0.1}}
\put(22.10,22.40){\circle*{0.05}}   \put(22.10,17.60){\circle*{0.1}}
\put(22.20,22.34){\circle*{0.05}}   \put(22.20,17.66){\circle*{0.1}}
\put(22.30,22.29){\circle*{0.05}}   \put(22.30,17.71){\circle*{0.1}}
\put(22.40,22.23){\circle*{0.05}}   \put(22.40,17.77){\circle*{0.1}}
\put(22.50,22.17){\circle*{0.05}}   \put(22.50,17.83){\circle*{0.1}}
\put(22.60,22.10){\circle*{0.05}}   \put(22.60,17.90){\circle*{0.1}}
\put(22.70,22.03){\circle*{0.05}}   \put(22.70,17.97){\circle*{0.1}}
\put(22.80,21.96){\circle*{0.05}}   \put(22.80,18.04){\circle*{0.1}}
\put(22.90,21.88){\circle*{0.05}}   \put(22.90,18.12){\circle*{0.1}}
\put(23.00,21.80){\circle*{0.05}}   \put(23.00,18.20){\circle*{0.1}}
\put(23.10,21.72){\circle*{0.05}}   \put(23.10,18.28){\circle*{0.1}}
\put(23.20,21.62){\circle*{0.05}}   \put(23.20,18.38){\circle*{0.1}}
\put(23.30,21.53){\circle*{0.05}}   \put(23.30,18.47){\circle*{0.1}}
\put(23.40,21.42){\circle*{0.05}}   \put(23.40,18.58){\circle*{0.1}}
\put(23.50,21.30){\circle*{0.05}}   \put(23.50,18.70){\circle*{0.1}}
\put(23.60,21.17){\circle*{0.05}}   \put(23.60,18.83){\circle*{0.1}}
\put(23.70,21.01){\circle*{0.05}}   \put(23.70,18.99){\circle*{0.1}}
\put(23.80,20.83){\circle*{0.05}}   \put(23.80,19.17){\circle*{0.1}}
\put(23.90,20.59){\circle*{0.05}}   \put(23.90,19.41){\circle*{0.1}}
\put(25,18){,}
\put(14,6){2nd}
\put(38,6){3rd}
\put(34,20){\circle*{0.6}}
\put(34,20){\line(1,-1){4}}
\put(34,20){\line(1,0){4}}
\put(41,20){\circle{14}}
\put(41,13){\line(-1,1){4}}
\put(41,13){\line(1,1){4}}
\put(41,13){\circle*{0.6}}
\put(41,20){\line(-1,0){3}}
\put(41,20){\line(1,0){4}}
\put(48,20){\line(-1,-1){4}}
\put(48,20){\line(-1,0){3}}
\put(48,20){\circle*{0.6}}
\put(50,18){,}
\put(60,6){4th (4a)}
\put(58,20){\circle*{0.6}}
\put(58,20){\line(1,1){4}}
\put(58,20){\line(1,-1){4}}
\put(65,20){\circle{14}}
\put(65,13){\circle*{0.6}}
\put(65,13){\line(-1,1){4}}
\put(65,13){\line(1,1){4}}
\put(65,27){\circle*{0.6}}
\put(65,27){\line(-1,-1){4}}
\put(65,27){\line(1,-1){4}}
\put(72,20){\circle*{0.6}}
\put(72,20){\line(-1,-1){4}}
\put(72,20){\line(-1,1){4}}
\put(74,18){,}
\put(84,6){4th (4b)}
\put(82,20){\circle*{0.6}}
\put(82,20){\line(1,1){4}}
\put(82,20){\line(1,0){4}}
\put(89,20){\circle{14}}
\put(89,20){\line(-1,0){4}}
\put(89,20){\line(0,1){4}}
\put(89,20){\line(0,-1){4}}
\put(89,20){\line(1,0){4}}
\put(89,13){\circle*{0.6}}
\put(89,13){\line(0,1){4}}
\put(89,13){\line(1,1){4}}
\put(89,27){\circle*{0.6}}
\put(89,27){\line(-1,-1){4}}
\put(89,27){\line(0,-1){5}}
\put(96,20){\circle*{0.6}}
\put(96,20){\line(-1,0){4}}
\put(96,20){\line(-1,-1){4}}
\put(101,18){and}
\put(118,20){\circle{14}}
\put(122,6){4th (4c)}
\put(118,13){\line(0,1){4}}
\put(118,13){\line(1,0){6}}
\put(118,20){\line(0,1){4}}
\put(118,20){\line(0,-1){4}}
\put(118,27){\line(0,-1){4}}
\put(118,27){\line(1,0){6}}
\put(127.5,13){\line(1,0){4}}
\put(127.5,13){\line(-1,0){4}}
\put(127.5,27){\line(1,0){4}}
\put(127.5,27){\line(-1,0){4}}
\put(137,20){\circle{14}}
\put(137,13){\line(0,1){4}}
\put(137,13){\line(-1,0){6}}
\put(137,20){\line(0,1){4}}
\put(137,20){\line(0,-1){4}}
\put(137,27){\line(0,-1){4}}
\put(137,27){\line(-1,0){6}}
\put(146,18){.}
\end{picture}   \ \ \ \ \
For these five diagrams from the left to the right, their symmetry factors
(the number of topologically-equivalent diagrams appearing in the expansion)
are $N_2=4!$, $N_3={\frac {3!}{3\cdot 2}}\cdot (2\cdot C_4^2)^3$,
$N_{4a}={\frac {4!}{4\cdot 2}}\cdot (2\cdot C_4^2)^4$, $N_{4b}={\frac {4!}
{4\cdot 2}}\cdot (C_4^2\cdot 2\cdot C_4^2)^2\cdot 2^4$ and $N_{4c}=
{\frac {4!}{4\cdot 2}}\cdot (C_4^3\cdot 3!\cdot C_4^3)^2\cdot 2$,
respectively. Thus, one can easily write down the corrections $F^{(2)}$,
$F^{(3)}$ and $F^{(4)}$ according to the above diagrams and then calculate
them as
\begin{eqnarray}
F^{(2)}&=&-{\frac {1}{\beta}}{\frac {\lambda^2}{2!}}N_2\int_0^\beta
         d\tau_1 d\tau_2 G^4_{\tau_1\tau_2}  \nonumber  \\
       &=&-{\frac {3\lambda^2}{64m^4\Omega^5}}\sinh^{-4}({\frac {\beta\Omega}
       {2}})[6\beta\Omega+8\sinh(\beta\Omega)+\sinh(2\beta\Omega)] \;,
\end{eqnarray}
\begin{eqnarray}
F^{(3)}&=&{\frac {1}{\beta}}{\frac {\lambda^3}{3!}}N_3\int_0^\beta
         d\tau_1 d\tau_2d\tau_3 G^2_{\tau_1\tau_2}G^2_{\tau_2\tau_3}
         G^2_{\tau_3\tau_1}  \nonumber  \\
       &=&{\frac {9\lambda^3}{512m^6\Omega^8}}\sinh^{-6}({\frac {\beta\Omega}
       {2}})\{-48+32\beta^2\Omega^2+[-3+8\beta^2\Omega^2]\cosh(\beta\Omega)
       \nonumber   \\
       &&+48\cosh(2\beta\Omega)+3\cosh(3\beta\Omega)+108\beta\Omega
       \sinh(\beta\Omega)\} \;,
\end{eqnarray}
and
\begin{eqnarray}
F^{(4)}&=&-{\frac {1}{\beta}}{\frac {\lambda^4}{4!}}\int_0^\beta
         d\tau_1 d\tau_2d\tau_3\tau_4\{N_{4a} G^2_{\tau_1\tau_2}
         G^2_{\tau_2\tau_3} G^2_{\tau_3\tau_4}G^2_{\tau_4\tau_1}
         \nonumber  \\
       &&  N_{4b} G^2_{\tau_1\tau_2}G^2_{\tau_3\tau_4}
         G_{\tau_2\tau_3} G^2_{\tau_2\tau_4}
         G_{\tau_1\tau_3}G^2_{\tau_1\tau_4}+N_{4c} G^3_{\tau_1\tau_2}
         G^3_{\tau_3\tau_4} G_{\tau_2\tau_3}G_{\tau_4\tau_1}\}
         \nonumber  \\
       &=&-{\frac {3\lambda^4}{32768\beta m^8\Omega^{12}}}
           \sinh^{-8}({\frac {\beta\Omega}{2}})
       \{6291-181320\beta^2\Omega^2+25920\beta^4\Omega^4 \nonumber  \\
       &&+6[71+13156\beta^2\Omega^2+2688\beta^4\Omega^4]\cosh(\beta\Omega)
      \nonumber   \\    &&
      + 48[-134+2115\beta^2\Omega^2+6\beta^4\Omega^4]\cosh(2\beta\Omega)
       \nonumber   \\
       &&-432\cosh(3\beta\Omega)+864\beta^2\Omega^2\cosh(3\beta\Omega)+
       141\cosh(4\beta\Omega)+6\cosh(5\beta\Omega) \nonumber  \\
       &&-191394\beta\Omega\sinh(\beta\Omega)+129456\beta^3\Omega^3
       \sinh(\beta\Omega)+42568\beta\Omega\sinh(2\beta\Omega)  \nonumber  \\
       && +12672\beta^3\Omega^3\sinh(2\beta\Omega)+37750\beta\Omega
       \sinh(3\beta\Omega)+1600\beta\Omega\sinh(4\beta\Omega)
       \} \;.
\end{eqnarray}
These analytical expressions of Eqs.(24), (25) and (26) are the main results
in this section. In order to obtain them, we have had to handle the absolute
value symbol in the expression of $G_{\tau\tau'}$ (see Eq.(18)). It is
straightforward to calculate the integrals in $F^{(2)}$ and $F^{(3)}$ by
dividing the integration domains into $2!$ and $3!$ parts respectively. As for
$F^{(n)} (n\ge 4)$, multi-dimensional integration domain, which exceeds our
direct intuition, is involved. However, for any $n$-dimensional integration
domain, one can divide it into $n!$ sub-domains so that, for each sub-domain,
the relation $\tau_{i_1}\le \tau_{i_2}\le\tau_{i_3}\le\cdots \le\tau_{i_j}
\cdots\le\tau_{i_{n-3}}\le\tau_{i_{n-2}}\le\tau_{i_{n-1}}\le\tau_{i_n}$ holds.
Then, mimicking the calculation of $F^{(3)}$, one can find the following
equivalent relation
\begin{equation}
\int_0^\beta d\tau_1\cdots \tau_n \Leftrightarrow \sum_P
\int_0^\beta d\tau_{i_n}\int_0^{\tau_{i_n}} d\tau_{i_1}
\int_{\tau_{i_1}}^{\tau_{i_n}} d\tau_{i_2}
\int_{\tau_{i_2}}^{\tau_{i_n}} d\tau_{i_3}
\cdots \int_{\tau_{i_{n-3}}}^{\tau_{i_n}} d\tau_{i_{n-2}}
\int_{\tau_{i_{n-2}}}^{\tau_{i_n}} d\tau_{i_{n-1}} \;,
\end{equation}
where the letter ``P'' below the summation symbol means the summation is
carried out over all the $n!$ sub-domains. Eq.(27) allows one to obtain
Eq.(26) with the aid of the computer software Mathematica.

Using the above results, we can now readily calculate the free energy up to
the fourth order : $F_2=F_0+F^{(2)}$, $F_3=F_0+F^{(2)}+ F^{(3)}$ and
$F_4=F_0+F^{(2)}+F^{(3)}+F^{(4)}$. In the following, we will numerically
compare them with existing results to examine rediability of our scheme.

First, we compare our results with the exact results obtained from
Ref.~\cite{6}. Using $F=-T\ln[\sum_n e^{-E_n/T}]$ ($E_n$ is the $n$th
eigenenergy of Eq.(1)), letting $m=\omega=1$, and for $T<1$, one can
calculate the exact free energies from Table V in Ref.~\cite{6}. For this
case, we plot Fig.1 with $\lambda=1$. In Fig.1, the dotted, short-dashed,
medium-dashed, long-dashed and solid curves are the exact free energy
$F_{exa}$, $F_0$, $F_2$, $F_3$ and $F_4$, respectively. Fig.1 indicates that
: (i) when the temperature is near zero, $F_2$ and $F_3$ are very to close
$F_{exa} $, whereas $F_4$ is unbounded from below; (ii) when the temperature
is greater than $0.5$ or so, $F_2$ and $F_4$ provide substantial corrections
to $F_0$, and $F_4$ gives better than $F_2$ does, while $F_3$ is close to
$F_0$. Here, we note that the invalidity of $F_4$ at very low temperature is
not unexpected. Since the present scheme is basically the Taylor expansion of
the free energy, the smallness of the temperature prevents and competes with
the convergence process of the perturbation and finally wins over at the
fourth order.

Then, we can compare our results with the accurate free energies, $F_{accu}$
from the Okopi$\acute{n}$ska's optimized variational method \cite{10}. In
order to compare with Okopi$\acute{n}$ska's data, we used the definitions of
the dimensionless quantities in Ref.~\cite{10}, that is, $m=1$, $z={\frac {1}
{2}}\omega^2\lambda^{-{\frac {2}{3}}}$, $\Omega\lambda^{-{\frac {1}{3}}}\to
\Omega$, $T \lambda^{-{\frac {1}{3}}}\to T$ and $F_i \lambda^{-{\frac {1}
{3}}}\to F_i$ \footnote{In Ref.~\cite{10}, there is a typo on the rescaling
expression of $T$, and here it is corrected.}. For the case of $z=10$, which
corresponds to $\lambda=0.01118$ in the dimensionlized system of
Ref.~\cite{6}, we give the comparison in Table I \footnote{The data of the
accurate free energies were provided by Okopi$\acute{n}$ska, the author of
Ref.~\cite{10}.}. From this table, one can see that $F_4$ has a better
agreement with $F_{accu}$ than $F_2, F_3$ and $F_0$ except for $T=30$.

Thirdly, to show the improvement of $F_0$ by higher-order corrections, for the
range of $1<T<50$, we plotted the results in Fig.2 to compare $F_4$ with $F_0$
in the cases of $z=0.2,1,10,30$ and $50$. In Fig.2, we use the same curve type
to represent $F_4$ and $F_0$, and between the curves within the same type,
$F_4$ is always the lower. Also, Fig.2 shows that $F_4$ coincides almost with
$F_0$ for $z=30$ and $50$, and the differences between $F_0$ and $F_4$ are
quite large for both $z=1$ and $z=0.2$. From Fig.2, we learn that : (i) for a
given temperature, with the increase of $z$, $i.e.$, with weakening  the
coupling, the corrections of $F_4$ to $F_0$ get smaller; (ii) for a given $z$,
with the increase of $T$, the corrections of $F_4$ to $F_0$ become larger;
(iii) the quite large differences between $F_0$ and $F_4$ imply that our
scheme becomes invalid with decreasing $z$ or with increasing coupling
strength (for a fixed $\omega^2$). The third point is similar to the optimized
expansion \cite{9} (1994).

Finally, taking $m=\omega=1$, we compare our results with those obtained from
the cumulant expansions \cite{6} in Table II. In Table II, the free energies
$F_2$ and $F_3$ are ours, the free energies $F^{Kr}_1$ and $F^{Kr}_3$ are the
first- and third-order results from the cumulant expansion in Ref.~\cite{6}
($i.e.$, $F_1$ and $F_2$ of Table II in Ref.~\cite{6}). The temperatures in
this table is lower than $1$ and the corresponding $z$ is small. So we did
not include $F_4$ in the table due to its invalidity. This table indicates
that our $F_2$ and $F_3$ are nearer to the exact value than $F^{Kr}_1$, but is
not so good as $F^{Kr}_3$. This reflects that the convergency of our expansion
is not so fast as in the cumulant expansions in Ref.~\cite{6}. As for the
optimized expansions, Ref.~\cite{9} compared the free energy with the exact
results in the case of both $\omega=0$ ($i.e.$, the smallest $z$) and the
reduced temperature less than $1$, and demonstrated a fast convergence. Here,
in Fig.3, our results are compared at various orders with Fig.2 in
Ref.~\cite{9} (1994) \footnote{The data from the optimized expansions were
also provided by Okopi$\acute{n}$ska.}. In Fig.3, from the upper to lower
(for large values of $\beta$), the first, second, sixth and seventh curves are
the variational, the third-, the second- and the fourth-order
results of our scheme, respectively. The third and the fourth curves are the
second- and the third-order results of the optimized expansions, and the fifth
curve is the exact result. The variational result is just the first-order
result in the optimized expansions. From this figure, our third-order result
is not so good as the results in the optimized expansions, and our
second-order result is almost as good as the second-order results in the
optimized expansions. This figure simultaniously indicates the invalidity of
our fourth-order result. Thus, the optimized expansion has also a better
convergency than our results. Additionally, the optimized expansions approach
the exact result monotonously in orders, whereas our results oscillate
with orders. However, we want to emphasize that our sheme is not so
complicated as the optimized and cumulant expansions. The crucial difference
between ours and the optimized or the cumulant expansions is that our scheme
performs the variational procedure at the lowest order and, accordingly, the
parameter $\Omega$ is identical for all orders, whereas in both the optimized
and the cumulant expansions \cite{9,6}, the variational procedure are performed
at their truncated order, and, consequently, $\Omega$ at one order is different
from the next. It is this difference that gives rise to the simplicity and slow
convergency of our scheme and the fast convergency and complication of the
optimized or cumulant expansions.

\section{Conclusion and Discussion}
\label{5}

In this paper, we have generalized the scheme in Ref.~\cite{8} to a Bosonic
case, and taking the anharmonic oscillator, Eq.(1), as a laboratory, provided
a wide test of its efficiency. Our investigations demonstrate that the present
scheme can substantially improve the variational result even in the second
order and when the reduced temperature is greater than $0.5$ or
so and the reduced coupling parameter $\lambda$ is not strong (or $z$ is
small), the free energy for Eq.(1) up to the fourth order in our expansion
gives a good agreement with the accurate result. We also demonstrate that
for the free energy of Eq.(1), when the reduced temperature approach zero, or
the reduced coupling $\lambda$ is strong (or $z$ is small), the fourth-order
result is invalid. Thus, from our investigations here, one can see the
efficiency and limitations of our scheme. Here, we also note that the present
scheme is much simpler than the optimized and cumulant expansions,
albeit it is not so fast convergent as they are. We believe that a simple
scheme is often necessary and useful because the non-perturbative method beyond the
Gaussian approximation is extremely complicated in general. Additionally, we
gave the approximate free energy of the system, Eq.(1), for moderate
temperature range. Although the exact results for the moderate temperature has
existed in the literature \cite{10}, our results can be readily used, as a
basis of quantitative comparison, when some other approximate methods produce
the free energy at the same temperature range.

In general, it should be noted that the variational perturbation theory
yields an asymptotic rather than a convergent series \cite{14}, and
hence a concrete range of its validity in a specific problem has no
universality. As for any specific problem, the present
scheme should always be used with a judicial examination of the true physical
property. We believe that the present paper can provide a qualitative
reference for an application of our scheme. In particular, when a specific
problem is too complicated to treat beyond the Gaussian approximation with
other expansions, we believe that our scheme can provide a simple and viable
tool to treat it.

Finally, we want to point out that, although we treated the quantum-mechanical
anharmonic oscillator only in this paper, it is straightforward to apply our
method to finite temperture scalar field theory \cite{15}. Especially, when it
is generalized to the $\phi^6$ models \cite{17}, we expect that the simplicity
of the method still holds there.

\acknowledgments
Lu acknowledges A. Okopi$\acute{n}$ska for providing her accurate data, and
would like to thank H. S. Park for his help. This project was supported
by the Korea Research Foundation (99-005-D00011). Lu's work was also supported
in part by the National Natural Science Foundation of China under the grant
No. 19875034.

\figure{Fig.1 \ \ \
        For the case of $T<1$ and $\lambda=1$, $F_2,F_3$ and $F_4$ the free
     energies up to the second, the third and the fourth orders are compared
      with the variational result $F_0$ and the exact free energy. The exact
        results were calculated according to $F=-T\ln[\sum_n e^{-E_n/T}]$ and
      Table V in Ref.~\cite{6}. Here, $E_n$ represents the $n$th eigenenergy
        for the system Eq.(1). We took $m=\omega=1$. In this figure, when
    $T>0.6$, the curve for $F_3$ almost coincides with the curve for $F_0$.}
\figure{Fig.2 \ \ \
     For the case of $T>1$ and for several values of $z$, $F_4$ are compared
     with $F_0$. We use the same type of curves to represent $F_4$ and $F_0$,
        and the latter is always above the former. But for the cases of $z=30$
        and $50$, $F_4$ almost coincides with $F_0$, and for the case of
        $z=1.0$ and $0.2$, $F_4$ is quite lower than $F_0$.}
\figure{Fig.3 \ \ \ Our results $F_2$, $F_3$ and $F_4$ are compared with the
         the second and third-order results by the optimized expansions which
         were provided by Okopi$\acute{n}$ska (z=0). In this figure, from the
         upper to lower (for large values of $\beta$), the first, the second,
         the sixth and the seventh curves are the variational, the third-, the
         second- and the fourth-order results of our scheme, respectively. The
         third and the fourth curves are the second- and the third-order
         results of the optimized expansions respectively, and the fifth curve
         is the exact result.}

\begin{table}
\caption{Our results $F_2, F_3$ and $F_4$ are compared with the
         variational result $F_0$ and the accurate free energies $F_{accu}$
         provided by Okopi$\acute{n}$ska (z=10).}
\begin{tabular}{|c|c|c|c|c|c|}
T & $F_4$ & $F_{accu}$ & $F_0$ & $F_2$ & $F_3$\\
\tableline
 1. &    2.262259  &    2.26225951564  &    2.262452  &    2.2622504 &
         2.262261 \\
 2. &    2.063913  &    2.06391575514  &    2.064409  &    2.0638734 &
         2.063925 \\
 3. &    1.555676  &    1.55569718863  &    1.556991  &    1.5555342 &
         1.555747 \\
 4. &    0.7808495 &    0.780936961496 &    0.7836171 &    0.7805129 &
         0.7811028 \\
 5. &   -0.2099735 &   -0.209722583045 &   -0.2050294 &   -0.210593  &
        -0.2093154 \\
10. &   -7.37775   &   -7.37249823358  &   -7.348171  &   -7.3793287 &
        -7.367283 \\
20. &  -28.03925   &  -27.9670036469   &  -27.86147   &  -28.0074342 &
       -27.92105 \\
30. &  -53.50769   &  -53.2269143165   &  -52.99767   &  -53.3278138 &
       -53.08786 \\
\end{tabular}
\end{table}

\begin{table}
\caption{Our results $F_2$ and $F_3$ are compared with the
         variational result $F_0$, the first- and third-order result,
         $F^{Kr}_1$ and $F^{Kr}_3$ obtained by the cumulant expansions ( $F_2$
         and $F_3$ in Table II of Ref.[6]) and the exact free energies
         $F_{exa}$. In this table, $m=\omega=1$}
\begin{tabular}{|c|c|c|c|c|c|c|c|}
$\lambda$ & $\beta$ & $F_{0}$ & $F^{Kr}_1$ & $F_3$ & $F^{Kr}_3$& $F_{exa}$ &
              $F_{2}$                \\
\tableline
    1.0&  5.0&  0.812491&  0.81188&  0.807364&  0.803882&  0.803758&
                0.800767\\
    5.0&  5.0&  1.244312&  1.24353&  1.2355&  1.22494 &  1.22459 &
                1.216996\\
   50.0&  5.0&  2.54758&  2.54675&  2.529673&  2.50067 &  2.49971 &
                2.480384\\
  500.0& 10.0&  5.425756&  5.42536 &  5.387961&  5.32211&  5.3199  &
                5.276719\\
20000.0&  3.0& 18.50166& 18.5003 & 18.37314& 18.1449  & 18.137   &
               17.98822\\
\end{tabular}
\end{table}

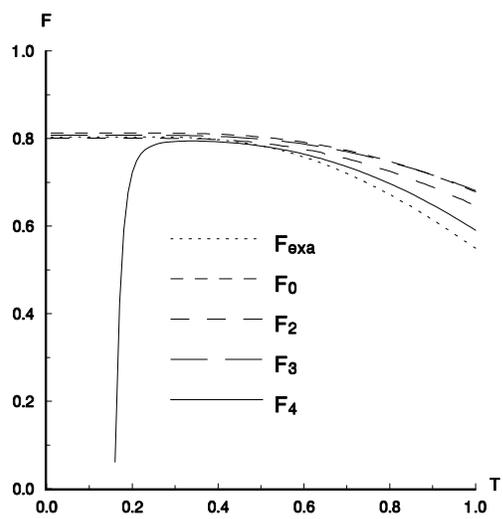
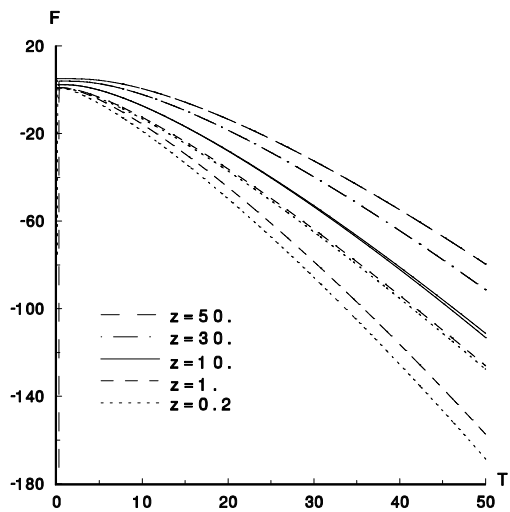
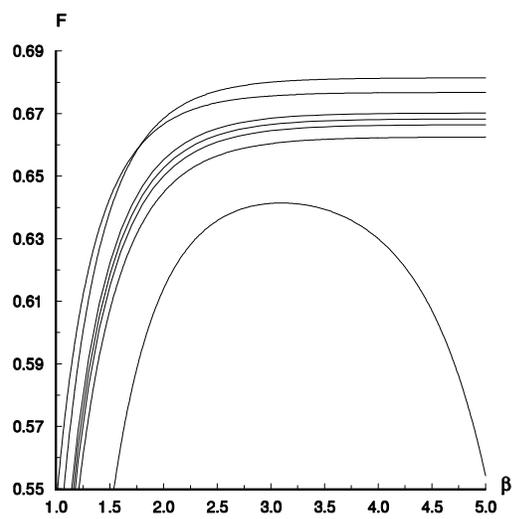
\end{document}